\input{epsf}
\documentclass [aps,pre,twocolumn,showpacs]{revtex4}
\usepackage[final]{graphics}
\usepackage{amssymb}
\usepackage{amsfonts}
\usepackage{epsfig}

\begin{document}

\title{Probability distribution of the order parameter}

\author{P. H. L. Martins$^{1,2}$\footnote{Email: pmartins@uft.edu.br} and 
J. A. Plascak$^{2}$\footnote{Email: pla@fisica.ufmg.br} }

\affiliation{$^1$Universidade Federal do Tocantins, 
Caixa Postal 111, 77001-970, Palmas, TO - Brazil \\ 
$^2$Departamento de F\'{\i}sica, Instituto de Ci\^encias Exatas, 
Universidade Federal de Minas Gerais, Caixa Postal 702, 30123-970, 
Belo Horizonte, MG - Brazil }

\date{\today}

\begin{abstract}

The probability distribution of the order parameter is exploited
in order to obtain the criticality of magnetic systems. 
Monte Carlo simulations have been employed by using single
spin flip Metropolis algorithm aided by finite-size scaling and
histogram reweighting techniques. A method is  proposed  
to obtain this probability distribution even when the transition temperature 
of the model is unknown. A test is performed on the two-dimensional 
spin-1/2 and spin-1 Ising model 
and the results show that the present procedure can be quite efficient
and accurate to describe the criticality of the system.

\end{abstract}


\maketitle

\section{Introduction}

\begin{figure}[htb]
\includegraphics[clip,angle=0,width=6.5cm]{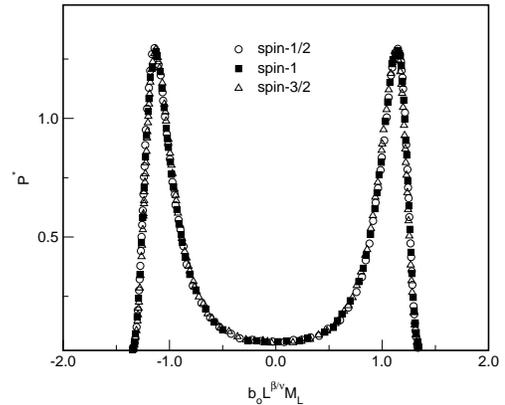}
\caption{\label{spins1-3}Scaling function $P^*(\tilde{m})$ 
for the two-dimensional 
spin-1/2, spin-1, and spin-3/2 Ising model on square lattices with $L=32$. 
Simulations were performed at the exact $T_c$ for spin-1/2, at $T_c = 1.6935$ 
for spin-1, according series expansions \cite{Adler} and 
Monte Carlo simulations 
\cite{Pla02}, and at $T_c=3.28794$ for the spin-3/2 model \cite{Pla02}.
The error bars are smaller than the symbol sizes. After Ref. \cite{Pla02}}
\end{figure}

The order parameter distribution function has been proved to be a 
powerful tool for studying not only magnetic systems 
\cite{Binder,Bruce,Nicolaides,Plascak,Tsypin}, 
but also the liquid-gas critical point
\cite{Wilding}, the critical point in the unified theory of weak and
electromagnetic interactions\cite{Rummuk}, and the critical point in quantum 
chromodynamics\cite{Alex}. For the specific case of magnetic
systems the order parameter  
can be chosen as the magnetization per spin, namely 
$m = \frac{1}{N} \sum_{i=1}^{N} S_i$, where $N$ is the total 
number of spins and $S_i$ is the spin at site $i$. In finite-size 
systems, the magnetization $m$ is a fluctuating quantity, 
characterized by the probability distribution $P(m)$ 
\cite{Binder,Bruce}. In Ising-like models undergoing a second-order
phase transition it is known that 
at temperatures lower than the critical 
temperature $T_c$, the distribution $P(m)$ has a double peak, 
centered at the spontaneous magnetizations $+m$ and $-m$. At temperatures 
greater than $T_c$, $P(m)$ has a single peak at zero magnetization, and 
exactly at $T_c$ a double-peak shape is observed\cite{Binder}.
Analogously to the usual finite-size scaling assumptions \cite{Fisher71}, 
one then expects that, for a large finite system of linear 
dimension $L$ at the critical point, $P(m)$ takes the form
\begin{equation}
P(m) = b P^* (\tilde{m}),
\label{pm}
\end{equation}
where $b = b_0 L^{\beta/\nu}$, $\beta$ and $\nu$ are critical exponents, 
$\tilde{m}= bm$, $b_0$ is a constant, and $P^*(\tilde{m})$ is a universal 
scaling function, normalized to unit norm and unit variance. 
Scaling functions, 
such as $P^*$, are characteristic of the corresponding universality class.  
Systems belonging to the same universality class share the same 
scaling functions. Thus, from the precise knowledge of $P^*(\tilde{m})$ 
one can characterize critical points and also identify universality 
classes. This is what has been done so far in the literature, with
the distribution for the spin-1/2 Ising model being the standard $P^*$ 
function\cite{Plascak,Wilding} for this universality class.
For instance, it is shown in Fig. \ref{spins1-3} the normalized 
distribution $P^*(\tilde{m})$ for the two-dimensional spin-1/2, spin-1, 
and spin-3/2 Ising model at criticality. Simulations have been done
on square lattices with $L=32$ 
at the exact $T_c$ for spin-1/2, at $T_c = 1.6935$ for spin-1, 
according series expansions \cite{Adler} and Monte Carlo simulations 
\cite{Pla02}, and at $T_c=3.28794$ for the spin-3/2 model \cite{Pla02}. 
The universal aspect of these systems can be easily noted. 
%
%%%%%%%%%%%%%%%%%%%%%%%% FIGURE 1 %%%%%%%%%%%%%%%%%%%%%%%%%%%%%%%%%%%%%%%%%%%%
%
%
%%%%%%%%%%%%%%%%%%%%%%%%%%%%%%%%%%%%%%%%%%%%%%%%%%%%%%%%%%%%%%%%%%%%%%%%%%%%%%%
%

Monte Carlo simulations seem to be the most effective method to
obtain results as those shown in Fig.\ref{spins1-3}, where the probability 
distribution $P(m)$ 
corresponds to the fraction of the total number of realizations in 
which the system magnetization is $m$, i.e., 
\begin{equation}
P(m) = \frac {N_m} {{\cal N}_{MCS}},
\label{pm2}
\end{equation}
where $N_m$ is the number of times that magnetization $m$ appears and 
${\cal N}_{MCS}$ is the total number of Monte Carlo steps. To compute the 
normalized distribution $P^*(\tilde{m})$ via Eq. (\ref{pm}) one has to 
evaluate the pre-factor $b$. This can be easily done by noting that 
$b = 1/\sigma$, where $\sigma$ is the square root of the magnetization 
variance ($\sigma^2 = \langle m^2 \rangle - \langle m \rangle^2$). 
Thus, one obtains the universal function $P^*(\tilde{m})$ by simply rescaling 
the magnetization and by using Eq. (\ref{pm}). 

In general, the probability distribution is used for studying models in 
which the critical temperature or even the distribution function is exactly 
(or high-precisely) known. That is in fact what has been done in the study
of several systems. When this distribution, as well as  the critical 
temperature and critical exponents, are not known, one can of course 
do first a 
canonical simulation in order to get the critical values (universal and
non universal) and compute, afterwards, the desired distribution.
The present approach is different from this conventional one in the sense
that it does use the order parameter distribution itself in order to
obtain the criticality of the system. 
The procedure, as well as the results obtained for the spin-1/2 and spin-1
Ising model, are discussed in  section II and
the conclusions are presented in the final section.

\section{Approach and results}

We have performed extensive Monte Carlo simulations (up to $10^7 - 10^8$ 
Monte Carlo steps per spin after $2.0 - 5.0 \times 10^4$ steps for 
thermalization) on square $L \times L$ lattices with periodic boundary 
conditions for systems of length $12 \le L \le 64$. For a given $L$, the 
simulation ran at a fixed temperature, evolving according the standard 
Metropolis algorithm. A histogram reweighting technique 
\cite{ron,Ferrenberg88} 
was used to obtain thermodynamic information in the vicinity of the 
simulated temperature. 

\begin{figure}[htb]
\includegraphics[clip,angle=0,width=11.0cm,height=9.5cm]{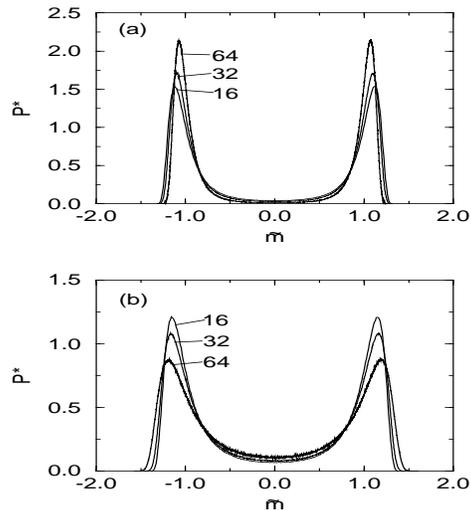}
\caption{\label{Fig2}Scaling function $P^*(\tilde{m})$ for the spin-1/2 
Ising model
with $L$ = 16, 32, and 64 at a fixed temperature $T$: 
(a) lower than $T_c$ ($T=2.2472$) and 
(b) greater than $T_c$ ($T=2.2831$). The error bars have been ommited
for clarity.}
\end{figure}

\begin{figure}[htb]
\includegraphics[clip,angle=0,width=6.5cm]{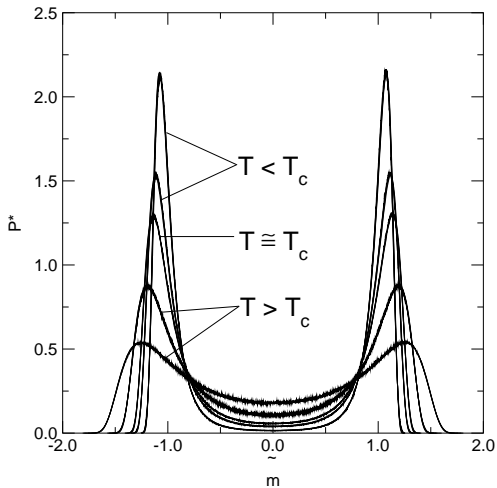}
\caption{\label{Fig3}Normalized distribution $P^*(\tilde{m})$ for systems 
with lattice sizes and temperatures shown in Table \ref{Tab1}. Each curve
is a supperposition of six different distributions taking with the
data from this table.}
\end{figure}

Let us first discuss the spin-1/2 Ising model. Figure \ref{Fig2} shows the 
distribution $P^*$ as a function of the normalized magnetization 
$\tilde{m}$ for temperatures different from the critical value $T_c$. 
As expected, one can 
see that for a temperature  lower than $T_c$ (Fig. \ref{Fig2}a), 
the maximum value of $P^*$ increases when the lattice size $L$ increases, 
while for a temperature greater than $T_c$, an increase of $L$ leads to a 
decrease of the corresponding peaks of $P^*$ (see Fig. \ref{Fig2}b). 
%
%%%%%%%%%%%%%%%%%%%%%%%%%% FIGURE 2 %%%%%%%%%%%%%%%%%%%%%%%%%%%%%%%%%%%%%%%%%%%
%%%%%%%%%%%%%%%%%%%%%%%%%%%%%%%%%%%%%%%%%%%%%%%%%%%%%%%%%%%%%%%%%%%%%%%%%%%%%%%
%
In other words, 
suppose we have a distribution function $P^*(\tilde{m})$ for a given $L$ 
(say for example, $L=16$) at a fixed temperature $T_{L=16}$. If 
$T_{L=16} < T_c$, a similar distribution will be obtained for a bigger 
lattice (e.g., $L=64$) at a different temperature $T_{L=64}$ such that 
$T_{L=16} < T_{L=64} < T_c$. Analogously, if $T_{L=16} > T_c$, we will 
have $T_{L=16} > T_{L=64} > T_c$. 
This fact suggests a mechanism to obtain the critical temperature, 
as well as the exponent $\nu$ and the universal distribution, for the system 
under study. Table \ref{Tab1} shows the temperatures of several lattice
%
%%%%%%%%%%%%%%%%%%%%%%%%% TABLE 1 %%%%%%%%%%%%%%%%%%%%%%%%%%%%%%%%%%%%%%%%%%%%%
%
\begin{table}[htb]
\caption{Temperature for different lattice sizes at which the distribution 
$P^*(\tilde{m})$ for $L=16-48$ is the same as that obtained for $L=64$ 
at the shown temperatures (spin-1/2). Error in 
parentheses affects the last digits. The second and third columns correspond 
to temperatures greater than the critical one, and the two following columns 
correspond to  temperatures lower than the critical one. 
The last column represents the data  when $P^*(\tilde{m})$ for
$L=64$ is obatined at $T_c$.} 
\begin{tabular}{cccccc}
\colrule
\colrule
\empty{Size} & \multicolumn{5}{c}{Temperature (in units of $J/k_B$)} \\ 
\colrule
16 & 2.3923(11) & 2.3272(8) & 2.2477(8) & 2.1901(10) & 2.27221(52) \\     
20 & 2.3666(8)  & 2.3154(8) & 2.2502(8) & 2.2036(10) & 2.27092(52) \\    
24 & 2.3502(8)  & 2.3073(5) & 2.2528(5) & 2.2134(7)  & 2.27015(52) \\
32 & 2.3288(5)  & 2.2973(5) & 2.2563(5) & 2.2262(5)  & 2.26963(26) \\
48 & 2.3089(5)  & 2.2878(5) & 2.2604(5) & 2.2399(5)  & 2.26937(26) \\
64 & 2.2989     & 2.2831    & 2.2624    & 2.2472     & 2.269184    \\
\colrule
\colrule
\end{tabular}
\label{Tab1}
\end{table}
%
%%%%%%%%%%%%%%%%%%%%%%%%%%%%%%%%%%%%%%%%%%%%%%%%%%%%%%%%%%%%%%%%%%%%%%%%%%%%%%%
% 
sizes we have used to obtain the distributions displayed in Fig. \ref{Fig3}. 
These temperatures were evaluated as follows. For $L=64$ and a given 
temperature, for instance $T=2.2989$ in Table \ref{Tab1}, we compute the 
corresponding probability distribution of the order parameter, which will be 
the ``reference'' distribution.  For other values of $L$, we search for the 
temperature $T_L$ 
which gives a distribution equivalent to the reference one. In this way,
we obtain the data shown in the second column of Table \ref{Tab1}. 
Taking a different reference distribution, 
obtained at a different temperature for $L=64$, we have another 
set of $T_L$, and so on. All the distributions so obtained are depicted in Fig.
\ref{Fig3}. It means that each curve in Fig. \ref{Fig3} is in fact
a superposition of six different distributions taking at the
temperatures given in Table \ref{Tab1}.
%
%%%%%%%%%%%%%%%%%%%%%%%% FIGURE 3 %%%%%%%%%%%%%%%%%%%%%%%%%%%%%%%%%%%%%%%%%%%%%
%
%
%%%%%%%%%%%%%%%%%%%%%%%%%%%%%%%%%%%%%%%%%%%%%%%%%%%%%%%%%%%%%%%%%%%%%%%%%%%%%%
%

%
%%%%%%%%%%%%%%%%%%%%%%%% FIGURE 4 %%%%%%%%%%%%%%%%%%%%%%%%%%%%%%%%%%%%%%%%%%%%
%
\begin{figure}[htb]
\includegraphics[clip,angle=0,width=12.0cm,height=9.0cm]{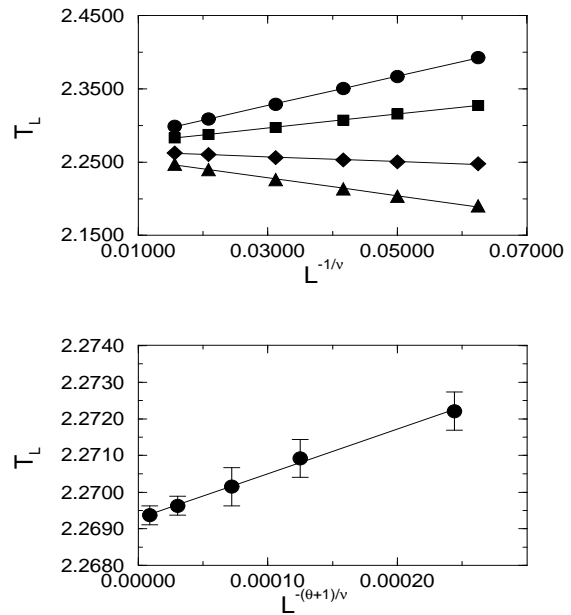}
\caption{\label{Fig4}(a) Temperature $T_L$ as a function of $L^{-1/\nu}$
with $\nu=1$. Different symbols correspond to different choices of the 
reference distribution (from top to bottom we have the results from
the second to fifth columns of Table \ref{Tab1}). Error bars are smaller
than the symbol sizes. 
(b) Temperature $T_L$ as a function of $L^{-(1+\theta)/\nu}$
with $\nu=1$ and $\theta=2$ taking the data of the last column of
Table \ref{Tab1}.
}
\end{figure}
%
%%%%%%%%%%%%%%%%%%%%%%%%%%%%%%%%%%%%%%%%%%%%%%%%%%%%%%%%%%%%%%%%%%%%%%%%%%%%%%
%

Since one expects that the difference 
$|T_L - T_c|$ scales as $L^{-1/\nu}$, where  $\nu$ is the correlation 
length critical exponent, a finite-size scaling analysis can be done to
estimate the critical values of the infinite system. 
In Fig. \ref{Fig4}a, we have a plot of
$T_L$ vs. $L^{-1/\nu}$, with $\nu=1$, using the values of the first 
five columns of Table \ref{Tab1}, which confirms the exact exponent 
$\nu=1$ and gives  
$T_c = 2.267(2)$. Another choice for this model is, of course, the
corresponding critical temperature. In this case
(very close to $T_c$, last column of Table \ref{Tab1}), however,
 it is known that 
$|T_L - T_c|$ scales as $L^{-(1+\theta)/\nu}$, where $\theta$ is the 
correction to scaling exponent \cite{Binder}.
In Fig. \ref{Fig4}b, we plot the estimates $T_L$ as a 
function of $L^{-(1+\theta)/\nu}$ with $\nu=1$ and
$\theta=2$ \cite{Calabrese00}. Linear regression gives $T_c=2.2693(1)$ for 
the infinite system, which is in fact quite close to the exact one.
 
%The procedure of using distributions very close to $T_c$ 
%have been largely employed in literature. Nevertheless, we show in this paper 
%that distributions at temperatures different from $T_c$ are also useful. We can 
%also observe two distinct forms of scaling, as $L^{-(1+\theta)/\nu}$ or 
%$L^{-1/\nu}$, if the system is or not close to the transition temperature. 

\begin{figure}[htb]
\includegraphics[clip,angle=0,width=6.5cm]{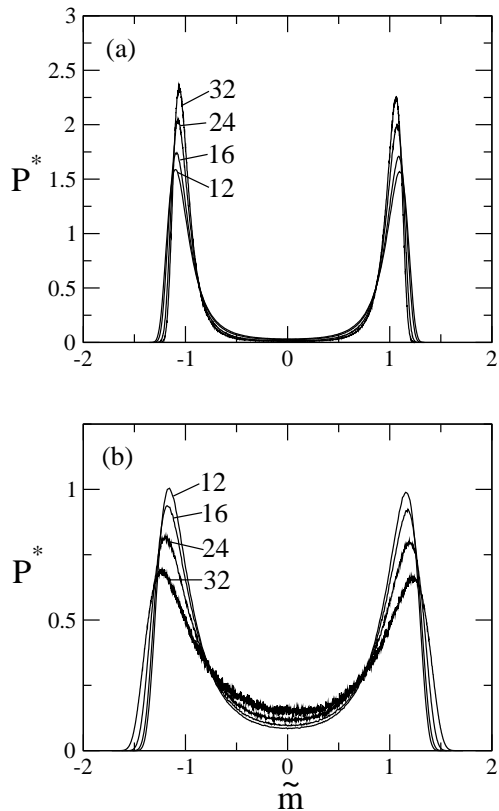}
\caption{\label{Fig-sp1a}Scaling function $P^*(\tilde{m})$ for the spin-1 model on 
square lattices with $L = 12, 16, 24$, and 32 at a fixed temperature $T$:
(a) lower than $T_c$ ($T=1.660$) and
(b) greater than $T_c$ ($T=1.720$).}
\end{figure}

In order to measure the applicability of the present mechanism for 
obtaining the transition temperature from 
non-universal distributions, we also study the spin-1 Ising model. 
To have an idea of 
the value of $T_c$, one just performs short simulations in a range of 
temperatures to 
check whether the probability distribution $P(m)$ has single or double peak. 
Then, one 
proceeds according to the same manner already discussed for the spin-1/2 
case. We fix 
the temperature and verify how the peaks of the distribution change if 
the lattice size $L$ 
increases. Figure \ref{Fig-sp1a} shows the distributions obtained for 
lattice sizes 
$L = 12$, 16, 24, and 32 at two different temperatures: $T = 1.660$ and 
$T=1.720$. 
In the former case an increasing lattice size leads to increasing peaks, 
while in the latter case 
the height of the peaks decreases when the lattice becomes larger. Thus, 
we conclude 
that the transition temperature is between 1.660 and 1.720, and hence we 
perform longer simulations in this temperature range.

%
%%%%%%%%%%%%%%%%%%%%%%%%%%% FIGURE 5 %%%%%%%%%%%%%%%%%%%%%%%%%%%%%%%%%%%%%%%%%%
%
%
%%%%%%%%%%%%%%%%%%%%%%%%%%%%%%%%%%%%%%%%%%%%%%%%%%%%%%%%%%%%%%%%%%%%%%%%%%%%%%%
%

\begin{figure}[tb]
\includegraphics[clip,angle=0,width=6.5cm]{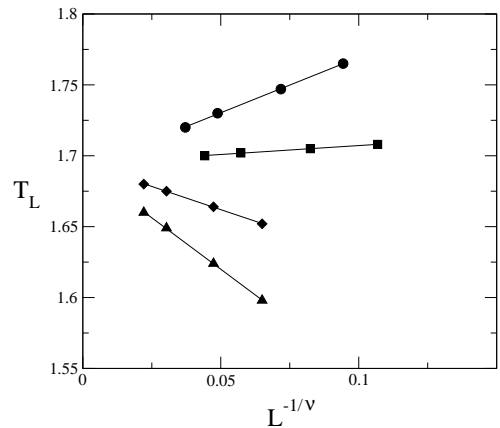}
\caption{\label{Fig-sp1b}Temperature $T_L$ as a function of $L^{-1/\nu}$ (spin-1). 
Different symbols correspond to different choices of the reference distribution 
(see Table \ref{Tab-sp1a}). The values of $\nu$ that give the best linear fit
were (from top 
to bottom): $\nu = 0.9(1)$, $\nu = 0.9(1)$, $\nu = 1.1(1)$, and $\nu = 1.1(1)$.
Error bars are smaller than the symbol sizes. 
} 
\end{figure}

\begin{figure}[htb]
\includegraphics[clip,angle=0,width=6.5cm]{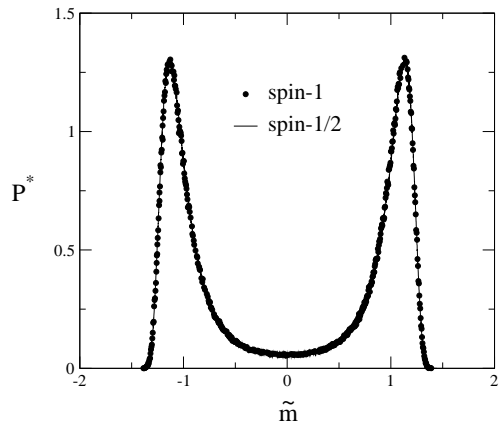}
\caption{\label{Fig-sp1c}Universal function $P^*(\tilde{m})$ for lattice size 
$L = 64$ at temperatures obtained in this work: $T=2.2693$ for spin-1/2 and 
$T=1.6933$ for spin-1.
} 
\end{figure}

The procedure now is the same as that we have done for the spin-1/2 model. 
We fix the 
temperature and compute $P^*(\tilde{m})$ for the lattice with $L=32$ (reference 
distribution). For a different $L$, we search for the temperature that gives a 
distribution equal to the reference one. Table \ref{Tab-sp1a} shows the 
temperatures so obtained. For each set of temperatures (which corresponds 
to each 
column of Table \ref{Tab-sp1a}), we plot $T_L$ vs. $L^{-1/\nu}$ and vary the 
exponent 
$\nu$ until we get a straight line. Thus, each column of Table \ref{Tab-sp1a} 
gives an 
independent estimate of $\nu$ and also of the critical temperature $T_c$. 
Figure 
\ref{Fig-sp1b} illustrates this procedure. By taking the mean value of these 
quantities, one obtains $\nu = 1.0(1)$ and $T_c=1.6933(16)$, where the
latter agrees well with the value $T_c=1.6935(10)$ \cite{Adler}. 
After we have evaluated $T_c$, we ran 
a longer simulation  on a larger lattice to determine, by this way, the 
universal distribution $P^*(\tilde{m})$. Figure \ref{Fig-sp1c} shows the 
distribution 
$P^*(\tilde{m})$ on a $L=64$ lattice for spin-1/2 and spin-1 models, and 
confirms  
the fact that both systems belong to the same universality class, as 
already expected. 

%
%%%%%%%%%%%%%%%%%%%%%%%%% TABLE 2 %%%%%%%%%%%%%%%%%%%%%%%%%%%%%%%%%%%%%%%%%%%%%%%
%
\begin{table}[htb]
\caption{Temperature for different lattice sizes at which the distribution 
$P^*(\tilde{m})$ for $L=12-24$ is the same as that obtained for $L=32$ 
at the shown temperatures (spin-1). Error in 
parentheses affects the last digits. The second and third columns correspond 
to temperatures greater than the critical one, and the two following columns 
correspond to  temperatures lower than the critical one.} 
\begin{tabular}{ccccc}
\colrule
\colrule
\empty{Size} & \multicolumn{4}{c}{Temperature (in units of $J/k_B$)} \\ 
\colrule
12 & 1.598(4) & 1.652(2) & 1.708(2) & 1.765(1) \\     
16 & 1.624(2) & 1.664(2) & 1.705(2) & 1.747(1) \\    
24 & 1.649(2) & 1.675(1) & 1.702(1) & 1.730(1) \\
32 & 1.660(2) & 1.680(1) & 1.700(1) & 1.720(1) \\
\colrule
\colrule
\end{tabular}
\label{Tab-sp1a}
\end{table}
%
%%%%%%%%%%%%%%%%%%%%%%%%%%%%%%%%%%%%%%%%%%%%%%%%%%%%%%%%%%%%%%%%%%%%%%%%%%%%%%%%
%
%
%%%%%%%%%%%%%%%%%%%%%%%%%% FIGURE 6 %%%%%%%%%%%%%%%%%%%%%%%%%%%%%%%%%%%%%%%%
%
%
%%%%%%%%%%%%%%%%%%%%%%%%%%%%%%%%%%%%%%%%%%%%%%%%%%%%%%%%%%%%%%%%%%%%%%%%%%%%%%%
%
%
%%%%%%%%%%%%%%%%%%%%%%%%%%%% FIGURE 7 %%%%%%%%%%%%%%%%%%%%%%%%%%%%%%%%%%%%%
%
%
%%%%%%%%%%%%%%%%%%%%%%%%%%%%%%%%%%%%%%%%%%%%%%%%%%%%%%%%%%%%%%%%%%%%%%%%%%%%%%%
%

%Concerning the correction to scaling exponent $\theta$, we did not evaluate it for 
%the spin-1 model. In the critical region, the statistical errors are large. 
%This makes difficult to choose different temperatures for different lattice sizes. 
%For example, at $T=1.693$, the distributions for $L=24$ and $L=32$ are equivalent, 
%within the errors. Thus, the evaluation of $\theta$ for this model needs longer 
%simulations. 

\section{Conclusions}

The present approach, using just the order parameter distribution, seems to
be a robust way to obtain the criticality of magnetic systems, regarding
its universal and non-universal aspects. There is a clear distinction between
the finite-size behavior of $P^*$  close to the
critical temperature (scaling with $L^{-(1+\theta)/\nu}$) or away from it
(scaling with $L^{-1/\nu}$), as depicted in Figs. \ref{Fig4} and \ref{Fig-sp1b}.
It also seems, at first sight, that there is a limitation regarding the size
of the lattices which could be considered.
For instance, we have used here  lattice sizes which are
smaller than that from the reference distribution.
Nevertheless, this limitation is not so drastic since in the spin-1 model we
considered the reference distribution for $L=32$ and with lattices smaller
than this value the results prove quite accurate. One can, of course, 
consider
lattices larger than that of the reference distribution. We feel, however,
that reweighting the distribution is more easily  done for  smaller systems.
Application of the present procedure to other models (pure and random),
as well as to multicritical behavior, will be very welcome; some are 
now in progress.

\acknowledgments

We would like to thank  J. G. Moreira and R. Dickman
for fruitful discussions. Financial support from the Brazilian agencies
CNPq, CAPES, FAPEMIG and
CIAM-02 49.0101/03-8 (CNPq) are gratefully acknowledged.


\begin{thebibliography}{99}

\bibitem{Binder}
K. Binder,
Z. Phys. B {\bf 43}, 119 (1981).

\bibitem{Bruce}
A.D. Bruce,
J. Phys. C {\bf 14}, 3667 (1981).

\bibitem{Nicolaides}
D. Nicolaides and A.D. Bruce,
J. Phys. A {\bf 21}, 233 (1988).

\bibitem{Plascak}
J.A. Plascak and D.P. Landau,
Phys. Rev. E {\bf 67} 015103 (R) (2003).

\bibitem{Tsypin}
M.M. Tsypin and H.W.J. Bl\"ote,
Phys. Rev. E {\bf 62}, 73 (2000).

\bibitem{Wilding} A. D. Bruce and N. B. Wilding, Phys. Rev. Lett.
 {\bf 68}, 193 (1992); N. B. Wilding and A. D. Bruce, J. Phys.: Condens.
 Matter {\bf 4}, 3087 (1992); N. B. Wilding, Phys. Rev. E {\bf 52},
 602 (1995).

\bibitem{Rummuk} K. Rummukainen, M. Tsypin, K. Kajantie, M. Laine, and
 M. Shaposhnikov, Nucl. Phys. B {\bf 532}, 283 (1998).

\bibitem{Alex} C. Alexandrou, A. Borici, A. Feo, P. de Forcrand,
A. Galli, F. Jegerlehner, and T. Takaishi, Phys. Rev. D {\bf 60},
034 504 (1999).


\bibitem{Fisher71}
M.E. Fisher in 
{\it Critical Phenomena}, 
edited by M.S. Green (Academic, New York, 1971).

\bibitem{Adler}
J. Adler and I.G. Enting, 
J. Phys. A {\bf 17}, L275 (1984).

\bibitem{Pla02}
J.A. Plascak, A.M. Ferrenberg, and D. P. Landau,
Phys. Rev. E {\bf 65}, 066702 (2002).

\bibitem{ron}
R. Dickman and W.C. Schieve, J. Physique {\bf 45}, 1727 (1984).

\bibitem{Ferrenberg88}
A.M. Ferrenberg and R.H. Swendsen,
Phys. Rev. Lett. {\bf 61}, 2635 (1988).

\bibitem{Calabrese00}
P. Calabrese, M. Caselle, A. Celi, A. Pelisseto, and E. Vicari,
J. Phys. A {\bf 33}, 8155 (2000);
J. Salas and A. Sokal, J. Stat. Phys. A {\bf 98}, 551 (2000);
S.L.A. de Queiroz, J. Phys. A {\bf 33}, 721 (2000).

\end{thebibliography}
\end{document}